\documentstyle[12pt]{article}

\begin{document}

\begin{titlepage}
\begin{center}
\today  \hfill  LBL-37478 \\
\hfill UCB-PTH-95/24 \\
\vskip .5 in

{\LARGE Large-N Baryons, Chiral Loops, and \\
   the Emergence of the Constituent Quark}
\footnote{This work was supported in part by the Director, Office of Energy
Research, Office of High Energy and Nuclear Physics, Division of High Energy
Physics of the U.S. Department of Energy under Contract DE-AC03-76SF00098.}

\vskip .3in

{\large Gregory L. Keaton}
\footnote{Present address:  Physics Department, Viginia Polytechnic Institute
and State University, Blacksburg, VA  24061. email:  keaton@mustang.phys.vt.edu}

\vskip .3in

{\it Department of Physics, University of California, Berkeley \\
and \\
Theoretical Physics Group, Lawrence Berkeley National Laboratory \\
Berkeley, CA  94720}
\end{center}

\vskip .5in

\begin{abstract}
Meson loop corrections to baryon axial currents are computed
in the $1/N$ expansion.  It is already known that the one-loop corrections are
suppressed by a factor $1/N$; here it is shown that 
the two-loop
corrections are suppressed by $1/N^2$.
To leading order, 
these corrections are exactly what would be calculated in the constituent
quark model.  Some applications are discussed.
\end{abstract}

\end{titlepage}

\newpage
\renewcommand{\thepage}{\arabic{page}}
\setcounter{page}{1}

\section{Introduction}

How are the baryons' properties renormalized by pion loops?  This classic
question gains renewed interest with the advent of each new
calculational technique.

Pion loop corrections to baryon properties have been studied using the
non-linear sigma model with derivative couplings \cite{old}.  Later, Jenkins and
Manohar \cite{jm1,jm2,jenkins} simplified the problem by invoking the heavy
baryon approximation \cite{bigm}.  Using a Lagrangian that included the baryon
octet and Goldstone boson octet, they found that the one-loop correction to the
baryon axial current was large---as much as 100\% of the tree-level value.
However, if the baryon decuplet is also included, the total one-loop 
corrections
are smaller, on the order of 30\% of the tree-level value.  That is, the loops
involving decuplet states tend to cancel the loops involving only octet states.
This was good news for perturbation theory, but it left unanswered the question,
``What is the loop expansion parameter?''  A seemingly coincidental cancellation
of large corrections did not leave behind any obvious parameter that could
justify, for example, the belief that the two-loop corrections should be any
smaller than the one-loop corrections.

This question can be addressed within the framework of large-N techniques for
baryons \cite{witten,gervais}, which have recently been rediscovered and
greatly expanded
\cite{sd1,sd2,djm1,harvard,lbl1,lbl2,lbl3,djm2}.  
One of the results is that the baryon coupling to the axial
current is on the order of the number of colors,
\begin{equation}
g_A \sim N
\label{ga}
\end{equation}
This also means that the baryon-pion coupling is $\sim g_A k_{\mu}/f \sim 
\sqrt{N}$.
However, the renormalization graph of Fig. 1 (a) gives a contribution of order
$N^2$ to $g_A$, which if taken alone would violate Eq. (\ref{ga}) and doom
perturbation theory (since the one-loop contribution would be much larger than
the tree-level value).  But there is another graph that must not be forgotten,
the wavefunction renormalization of Fig. 1 (b).  When both of these diagrams are
included, it is found \cite{sd1,djm1} 
that the leading order behaviour cancels,
and the total one-loop correction is ${\cal O}(1)$, or $1/N$ times the
tree-level value.  The one-loop corrections are therefore small in the 
$1/N$ expansion and chiral perturbation theory seems to be valid.

As encouraging as this result is, it leaves some questions open.  If the
one-loop results are to be fitted to the data and believed, one should show 
that
the two-loop contribution is small compared to the one-loop result.  
This is not
obvious since, for example, the diagram of Fig. 2(a) is of order $N^3$ times the
one-loop correction.  
However, once again there are several diagrams to be added
together.  This paper shows that when all of the two-loop 
diagrams are taken
into account, the largest terms cancel, and the result is of order $1/N$ times
the one-loop contribution.  Evidently, the pion loop expansion parameter turns
out to be $1/N$.  Although this has been suspected before \cite{sd1,djm1}, 
it has not been previously demonstrated to two loops.

One result of this analysis is a demonstration that when the pion-baryon
vertex is taken to leading order in $1/N$, 
the chiral loop corrections are the same as in the chiral quark model.  
This is surprising because in that model \cite{georgimanohar,jm2} the quarks
are considered to be weakly bound, so that the pions interact with only one
quark at a time, and the other quarks are spectators.  In the present $1/N$
analysis, however, the quarks are so tightly bound that the pion
interacts with the baryon as a whole, and there are no spectators. 
Nevertheless, when all of the diagrams are taken into account, many cross-terms
cancel, and the $1/N$ result is just what would have been calculated in the
chiral quark model. 
It had already
been noted that the constituent quark model fits the data as well as the usual
baryon-pion theory \cite{jm2}; the $1/N$ expansion sheds some light on why this
is so.

The next section presents the two-loop calculation; this is followed by a
discussion of possible applications of this formalism.

\section{Two-Loop Corrections}

What are the meson loop corrections to the baryon axial current?  The spatial
components of the axial current are written \cite{sd1,djm1,harvard}
\[
<{\cal B}' | \bar{q} \gamma^i \gamma_5 T^a q |{\cal B}>_{tree}
 = X^{ia}_{ B' B}
\]
where $T^a$ is a generator of the flavor group.  (The magnitude of $X^{ia}$ is
what was loosely called $g_A$ in the introduction.)  The axial current can be
expressed in the $1/N$ expansion using perturbative baryon states $|B>$
\cite{lbl1} (to be contrasted with physical states $|{\cal B}>$):
\[
|B> = B^{r_1 \alpha_1 \ldots r_N \alpha_N} a^{\dagger}_{r_1 \alpha_1}
 \ldots a^{\dagger}_{r_N \alpha_N} |0>
\]
Here $a^{\dagger}_{r\alpha}$ creates a quark with spin $\alpha$ and isospin $r$.
The quarks are totally antisymmetric with respect to color; their color
indices are suppressed, and the operators $a$ and $a^{\dagger}$ are
treated as bosonic, rather than fermionic \cite{lbl1}.  Then the axial current
can be written in a $1/N$ expansion
\begin{equation}
X^{ia}_{B'B} = g<B'|G^{ia}|B> + \frac{h}{N} <B'|H^{ia}|B> + \ldots
\label{x}
\end{equation}
where $g$ and $h$ are constants of order 1 and $G^{ia}$ and $H^{ia}$ are the
operators \cite{lbl1,lbl2,djm2}
\[
G^{ia} = a^{\dagger}_{r \alpha} T^a_{rs} \sigma^i_{\alpha \beta} 
  a_{s\beta}
\]
\[
H^{ia} = (a^{\dagger}_{r \alpha} T^a_{rs} a_{s \alpha})
  (a^{\dagger}_{t \beta} \sigma^i_{\beta \gamma} a_{t \gamma})
\]

The mesons are coupled derivatively to the baryon axial current.  Some of the
Feynman rules for the pion-baryon interactions are given in Fig. 3.  The
baryons are treated within the heavy fermion approximation \cite{bigm,jm1,jm2},
and the calculations are performed in the baryon's rest frame.  The meson
propagator uses the mass matrix $m^2_{ab}$, a diagonal matrix that gives the
masses of the pions, kaons, and eta under flavor symmetry breaking.  For the
$N$ power counting, it is important to keep in mind that the pion
decay constant $f \propto \sqrt{N}$.

Now look at the vertex renormalization.  The momentum integral for Fig. 1(a) is
\[
{\cal I}_{ab} = i \int \frac{d^4p}{(2\pi)^4} \: \frac{1}{p_0^2} \:
\frac{ \frac{1}{3} {\bf p}^2}{p^2 - m^2_{ab}}
\]
For Fig. 2(a) and Fig. 2(c), the integral is
\[
{\cal J}^{aa'bb'}_1 = - \int \frac{d^4p}{(2\pi)^4} \frac{d^4q}{(2\pi)^4}
\: \frac{1}{p_0^3}\left( \frac{1}{p_0 + q_0} - \frac{1}{q_0} \right)
\frac{ \frac{1}{3} {\bf p}^2}{p^2 - m^2_{aa'}} \:
\frac{ \frac{1}{3} {\bf q}^2}{q^2 - m^2_{bb'}}
\]
The inner loop includes a counter-term.  If this term (the $1/q_0$ appearing
above) is not included, the internal baryon acquires an additional mass, which
must then be transformed away by the heavy baryon transformation.  It is easier
to simply include the counter-term explicitly.  The mass differences between the
various baryons are proportional to $1/N$ and/or to the flavor symmetry
breaking, and will be ignored.

The integrals for Fig. 2(b),(d),(e), and (f) are (respectively)
\begin{eqnarray*}
{\cal J}_2^{aa'bb'}  & = & - \int \frac{d^4p}{(2\pi)^4} \frac{d^4q}{(2\pi)^4}
\: \frac{1}{p_0^2} \: \frac{1}{(p_0 + q_0)^2} \:
\frac{ \frac{1}{3} {\bf p}^2}{p^2 - m^2_{aa'}} \:
\frac{ \frac{1}{3} {\bf q}^2}{q^2 - m^2_{bb'}}  \\
{\cal K}_1^{aa'bb'}  & = & - \int \frac{d^4p}{(2\pi)^4} \frac{d^4q}{(2\pi)^4}
\: \frac{1}{p_0^2} \: \frac{1}{p_0 + q_0} \: \frac{1}{q_0} \:
\frac{ \frac{1}{3} {\bf p}^2}{p^2 - m^2_{aa'}} \:
\frac{ \frac{1}{3} {\bf q}^2}{q^2 - m^2_{bb'}}  \\
{\cal K}_2^{aa'bb'}  & = & - \int \frac{d^4p}{(2\pi)^4} \frac{d^4q}{(2\pi)^4}
\: \frac{1}{p_0} \: \frac{1}{(p_0 + q_0)^2} \: \frac{1}{q_0}
\frac{ \frac{1}{3} {\bf p}^2}{p^2 - m^2_{aa'}} \:
\frac{ \frac{1}{3} {\bf q}^2}{q^2 - m^2_{bb'}}  \\
{\cal K}_3^{aa'bb'}  & = & - \int \frac{d^4p}{(2\pi)^4} \frac{d^4q}{(2\pi)^4}
\: \frac{1}{p_0} \: \frac{1}{p_0 + q_0} \: \frac{1}{q_0^2} \:
\frac{ \frac{1}{3} {\bf p}^2}{p^2 - m^2_{aa'}} \:
\frac{ \frac{1}{3} {\bf q}^2}{q^2 - m^2_{bb'}}  = {\cal K}_1^{bb'aa'}  \\
\end{eqnarray*}
The vertex renormalization to two loops can then be written:
\begin{eqnarray}
V^{ia}_{B'B} & = & \left( X^{ia} 
+ \frac{1}{f^2} {\cal I}^{bb'}X^{jb'}X^{ia}X^{jb}
+ \frac{1}{f^4} {\cal J}_1^{bb'cc'}X^{jb}X^{ia}X^{kc}X^{kc'}X^{jb'} 
 \right. \nonumber \\
 & & \left.
+ \frac{1}{f^4} {\cal J}_2^{bb'cc'}X^{jb}X^{kc}X^{ia}X^{kc'}X^{jb'} 
+ \frac{1}{f^4} {\cal J}_1^{bb'cc'}X^{jb}X^{kc}X^{kc'}X^{ia}X^{jb'} 
\right. \nonumber \\
 & & \left.
+ \frac{1}{f^4} {\cal K}_1^{bb'cc'}X^{jb}X^{ia}X^{kc}X^{jb'}X^{kc'}
+ \frac{1}{f^4} {\cal K}_2^{bb'cc'}X^{jb}X^{kc}X^{ia}X^{jb'}X^{kc'} 
\right. \nonumber \\
 & & \left.
+ \frac{1}{f^4} {\cal K}_1^{cc'bb'}X^{jb}X^{kc}X^{jb'}X^{ia}X^{kc'} 
 \right)_{B'B} 
\label{vertex}
\end{eqnarray}
The operators $X^{ia}$ are treated as matrices with baryon indices; intermediate
baryon states are summed over.

The baryon wavefunction renormalization constant can be computed from the
diagrams of Fig. 1(b) and Fig. 4:
\begin{eqnarray}
\left( Z^{-1}_2 \right)_{B'B} & = & \left( 1 
 + \frac{1}{f^2} {\cal I}^{bb'} X^{jb} X^{jb'}
 + \frac{1}{f^4} (2 {\cal J}^{bb'cc'}_1 + {\cal J}^{bb'cc'}_2)
     X^{jb}X^{kc}X^{kc'}X^{jb'} \right. \nonumber \\
 & & \left. + \frac{1}{f^4} ({\cal K}^{bb'cc'}_1 + {\cal K}^{bb'cc'}_2
+ {\cal K}^{cc'bb'}_1 ) X^{jb}X^{kc}X^{jb'}X^{kc'} \right)_{B'B} 
\label{z2}
\end{eqnarray}
Finally, the renormalized axial current is
\begin{equation}
<{\cal B}'|\bar{q}\gamma^i \gamma_5 T^a q|{\cal B}>
 = \left( Z^{\frac{1}{2}}_2 \: V^{ia} \: Z^{\frac{1}{2}}_2 \right)_{B'B}
\label{ren}
\end{equation}

When Eqs. (\ref{vertex}) and (\ref{z2}) are substituted into Eq. (\ref{ren})
and the result is multiplied out to order $1/f^4$ (i.e. to two loops),
the terms do not simplify in any obvious way.  Some identities among the
integrals must be used,
\[
{\cal J}_2^{aa'bb'} + {\cal K}_2^{aa'bb'} = {\cal K}_1^{aa'bb'}
\]
\[
{\cal K}_1^{aa'bb'} + {\cal J}_1^{aa'bb'} = 0
\]
\[
{\cal K}_1^{aa'bb'} + {\cal K}_1^{bb'aa'} - {\cal I}^{aa'}{\cal I}^{bb'} = 0
\]
as well as the identity
\[
{\cal K}_1^{aa'bb'}[X^{ia}X^{ia'},X^{jb}X^{jb'}] = 0
\]
Then after a few pages of algebra, Eq. (\ref{ren}) becomes
\begin{eqnarray}
\, & \, & <{\cal B}'|\bar{q}\gamma^i \gamma_5 T^a q|{\cal B}>
 = X^{ia} + \frac{1}{2f^2}{\cal I}^{bb'} [X^{jb},[X^{ia},X^{jb'}]] \nonumber \\
 & & \;\;\;\;\; + \frac{1}{f^4}{\cal K}_{1}^{bb'cc'} \left\{
 \frac{1}{4} \left[ X^{jb},[[X^{kc},[X^{ia},X^{kc'}]],X^{jb'}] \right] 
 \right. \nonumber \\
 & & \;\;\;\;\;\;\;\;\;\; + \left.
\frac{1}{2} \left[ [X^{jb},X^{ia}],[X^{kc},[X^{jb'},X^{kc'}]] \right] 
 \right. \nonumber \\
 & & \;\;\;\;\;\;\;\;\;\; + \left. 
\frac{1}{4}\left[ X^{ia}, \left[ X^{jb}, [X^{kc},[X^{jb'},
 X^{kc'}]] \right] \right] \right\} \nonumber \\
 & & \;\;\;\;\; + \frac{1}{4f^4} {\cal K}^{bb'cc'}_2 \left[ [X^{jb},X^{kc}],
  [X^{ia},[X^{jb'},X^{kc'}]] \right] 
\label{result}
\end{eqnarray}
This is the main result of the paper.  From here it is possible to show that the
one-loop corrections to the axial current are suppressed by ${\cal O}(1/N)$
times the tree-level, and the two-loop contribution is suppressed by ${\cal
O}(1/N^2)$.

For example, suppose we take $X^{ia}$ to leading order in $1/N$,
\begin{equation}
X^{ia} = g<B'|G^{ia}|B>
\label{leadx}
\end{equation}
where $G^{ia} = a^{\dagger}_{r\alpha}T^a_{rs}\sigma^i_{\alpha \beta} 
a_{s\beta}$.
Since the operator $G^{ia}$ has one $a$ and one $a^{\dagger}$, it can count the
number of quarks in the baryon once, and can therefore 
be of order $N$. No accidental
cancellations occur, so the axial current is ${\cal O}(N)$ to tree level.
Now using Eq. (\ref{leadx}), the one-loop correction can be read off from Eq.
(\ref{result}):
\begin{equation}
\frac{g^3}{2f^2} {\cal I}^{bb'}<B'|[G^{jb},[G^{ia},G^{jb'}]]|B>
\label{oneloop}
\end{equation}
Each $G^{ia}$ has one $a$ and one $a^{\dagger}$, but each commutator eliminates
an $a$--$a^{\dagger}$ pair due to the identity 
$[a^{\dagger}_{a \alpha}V^{b\beta}_{a\alpha}a_{b\beta},
a^{\dagger}_{c\gamma}W^{d\delta}_{c\gamma}a_{d\delta}] = 
a^{\dagger}_{a\alpha}[V,W]^{b\beta}_{a\alpha}a_{b\beta}$.  The resultant
operator in Eq. (\ref{oneloop}) has only one $a$ and one $a^{\dagger}$, so the
matrix element is at most of order $N$.  Since $1/f^2 \sim 1/N$, the total
one-loop correction is ${\cal O}(1)$, or $1/N$ times the tree-level value.

The quadruple commutators, which give the two-loop corrections,
also eliminate all
but one $a$--$a^{\dagger}$ pair.  Therefore these commutators are also of 
${\cal O}(N)$, and when they are multiplied by a coefficient of $1/f^4$, the
result is ${\cal O}(1/N)$.  That is, the two-loop correction is $1/N^2$ times
the tree-level value.

This result has the following interpretation:  since the loop corrections
contain only one $a$ and one $a^{\dagger}$, the pion vertices and the current
operator all act on the same quark; the vertex and wavefunction
renormalization are carried out on each quark individually.  

The reason that this happens in the one-loop case can easily be demonstrated
graphically.  In Fig. 5(a), the ends of the pion line are attached to different
quark lines than the one on which the current acts.  In such a case, one of
the pion vertices can be commuted past the current operator, so that the pion
is emitted and absorbed before the current operator acts (see Fig. 5(b)). 
However, this diagram has already been taken into account by the wavefunction
renormalization, and so does not contribute.  Similarly, in the graph of 
Fig. 5(c), the pion line begins on a different quark line from the one on which
the current acts, but ends on the same line.  In this case, the first vertex
can be commuted past the current operator, resulting in Fig. 5(d).  Again, this
diagram is cancelled by the wavefunction renormalization.  So we reach the
conclusion stated above:  the only type of graph that needs to be considered
has both pion vertices acting on the same quark line as the current operator
(Fig. 5(e)).

This is exactly what is assumed to be true in the chiral quark model
\cite{georgimanohar} as developed in Ref. \cite{jm2}.  In that model the
leading vertex and wavefunction renormalizations act on one quark at a time,
and the quark propagator is that of a (fairly heavy) constituent quark.  One
problem with the chiral quark model, however, is that it is difficult to
understand why a free quark propagator can be used inside the proton.  Why
should not the bound state wavefunction be taken into account?  The large-N
approach offers a solution:  the quark inherits the heavy fermion propagator
from the baryon as a whole but manages to decouple from the other quarks thanks
to the cancellations that occur in the commutator structure of the loop
corrections.  The constituent quark emerges from a tangle of meson loops.

There are two points that complicate the above picture somewhat.
The first is that we have left out some diagrams.
Figs. 6 (a) - (e) also contribute \cite{jm1,jm2}; they are most easily 
calculated
using an effective Lagrangian \cite{lbl1,lbl2,lbl3}.  It turns out that these 
diagrams
follow the same pattern as above:  $l$-loop diagrams are suppressed by factors
of $(1/N)^l$, and the result is just what would have been expected from the
chiral quark model.  One difference of the these diagrams, however, is that they
are not necessarily suppressed by powers of the coupling 
constant $g$ of Eq.(\ref{leadx}).  For example,
Fig. 6(a) and Fig. 6(b) are both proportional to $g$ (rather than $g^3$ or
$g^5$).
This point does not affect the present discussion, but is important in the
next section.

The second complication is that Eq. (\ref{leadx}) is only an 
approximation to the
axial vertex.  When the vertex is expanded to the next order in $1/N$ (as
suggested by Eq. (\ref{x}) ), 
a new operator $H^{ia}$ is introduced.  $H^{ia}$ acts on
two quarks at a time, so the simplest constituent quark 
picture receives corrections\footnote{Actually such operators 
appear in the chiral quark model also;
rather than being suppressed by $1/N$, though, they are suppressed
\cite{georgimanohar} by a power of the wavefunction at the origin divided by the
constituent quark mass, $|\psi(0)|^{2/3}/m_c$. }.
The axial current is of ${\cal O}(N)$, and the leading corrections are of
${\cal O}(1)$.  To evaluate all the corrections to ${\cal O}(1)$, the operator
$H^{ia}$ must be included at tree level, as well as in the double commutator of
Eq. (\ref{result}).
(One might have thought that the loop
corrections involving the operator $H^{ia}$ would be smaller in $1/N$ 
than $H^{ia}$
itself, but this turns out not to be the case.)  Similarly, in order to
calculate loop corrections to ${\cal O}(1/N)$, 
$X^{ia}$ must be expanded to ${\cal O}(1/N)$ and then inserted into Eq.
(\ref{result}).  Nevertheless,
the identities of Ref. \cite{djm2} can be used to show that the one- and
two-loop corrections are still of ${\cal O}(1)$ and ${\cal O}(1/N)$
respectively.  In actual practice, Ref. \cite{lbl2} has estimated $h \approx
-0.1$, and so the operator $H^{ia}$ will be ignored below.

\section{Discussion}

How does all this formalism apply to the real world?  The double commutator 
in Eq. (\ref{result}), which
gives the one-loop correction, is of the order of the
number of flavors $N_F$, and the quadruple commutators giving the 
the two-loop corrections are of order $N_F^2$.  The momentum integrals should be
cut off at the chiral symmetry-breaking scale $\Lambda$.  Let the pion
decay constant $f$ be factorized to show clearly its N-dependence:
\[
f=\sqrt{N} \hat{f}
\]
where $\hat{f}$ is ${\cal O}(1)$.   
As mentioned previously, some diagrams of Fig. 6 are
not suppressed by powers of the axial coupling constants ($g$ and $h$ of Eq.
(\ref{x})).  Therefore the chiral loop expansion parameter is 
\[
\frac{N_F}{N} \: \frac{\Lambda^2}{16 \pi^2 \hat{f}^2}
\]
This parameter is not small in any estimation \cite{bluegeorgi}.  However, one
can adopt the following approach:  start with the bare coupling $g$ (or for
example $h$), assume that it can be renormalized to all orders in the flavor
symmetric limit ($m_{\eta} = m_K = m_{\pi} \approx 0$), resulting in the
renormalized constant $g_R$.  This new constant $g_R$ is to be
used in computations, and the effects of virtual pions can be computed
loop-by-loop, keeping only those terms that {\it violate} $SU(N_F)$ symmetry.
In this case all terms involving $\Lambda^2/16 \pi^2 \hat{f}^2$ are to be thrown
away, since their effects have already been included in $g_R$.  The new
loop expansion parameter then becomes
\[
\frac{N_F}{N} \: \frac{m_K^2}{16 \pi^2 \hat{f}^2} \log \frac{\Lambda^2}
{m_K^2}
\]
This procedure is equivalent to using dimensional regularization for all the
integrals of Eq. (\ref{result}).

Such a program has already been carried out by Ref. \cite{jm2}.  The one-loop
corrections to the baryon axial current were computed in the chiral quark model
using dimensional regularization.  This is exactly equivalent to a leading order
$1/N$ calculation.  The model fits the data well; a best fit is obtained for 
$g_R = 0.56$.

So far we have examined 
the corrections to the octet axial currents.  What
about the singlet current, the ``spin content'' of the baryon?
When computing the renormalization of the singlet current, fewer diagrams appear
than for the octet current:  Figs. 6 (a)--(c) do not exist for the singlet
current.  Therefore the one-loop contribution is suppressed by a factor
$g_R^2 (N_F/N)(\Lambda^2/16 \pi^2 \hat{f}^2)$
compared to the tree value.  The two-loop diagrams are suppressed by a factor of
$g_R^2(N_F/N)(\Lambda^2/16 \pi^2 \hat{f}^2)$
 (Figs. 2 and 4) or $(N_F/N)(m_K^2/16 \pi^2 \hat{f}^2)(\log \Lambda^2/m^2_K)$ 
(Figs. 6(d) and (e)) compared to the one-loop diagrams.  Therefore, the
loop expansion parameter $\epsilon$ for the singlet current is
\[
\epsilon = \max\left(g_R^2\frac{N_F}{N} \: \frac{\Lambda^2}{16 \pi^2 \hat{f}^2}
,\frac{N_F}{N} \: \frac{m_K^2}{16 \pi^2 \hat{f}^2} \log 
\frac{\Lambda^2}{m_K^2} \right)
\]
If $\epsilon$ is small enough, we do not have to go through the extra step of
first doing the flavor-symmetric renormalization and then returning to the
integrals using dimensional regularization.  Cutoff regularization can be used
from the outset.

Using this approach, and the leading $1/N$ baryon-pion vertex,
the spin content of the proton turns out to be
\begin{eqnarray}
 & & <p\uparrow|\bar{q}\gamma^3\gamma_5 q|p\uparrow> = 
g \left\{ 1 + \frac{g_R^2}{2f^2} {\cal I}^{bb'} <p\uparrow|
[G^{jb},[\sigma^3,G^{jb'}]]|p\uparrow> \right\} \nonumber \\
 & & \;\;\;\;\; = 
g \left\{1-g_R^2\left[ \frac{32}{9} \: \frac{\Lambda^2}{16 \pi^2 f^2} 
 - \frac{22}{9} \: \frac{m_K^2}{16 \pi^2 f^2}\:\log\frac{(2\Lambda)^2}
{m_K^2} \right] \right\}
\label{spin}
\end{eqnarray}
(Here I used $m^2_{\pi} = 0$ and $m_{\eta}^2 = \frac{4}{3}m^2_K$.)
This calculation assumes that $\Lambda$ is a physically meaningful cutoff (it is
the point where chiral symmetry is restored).  This assumption differs from
other chiral perturbation calculations \cite{old,jm1,jm2,jenkins,lbl3} where
the quadratically divergent pieces are discarded and only the logarithmically
divergent pieces are kept.  However, retention of the quadratically divergent
pieces is well within the spirit of the effective Lagrangian
\cite{georgimanohar}.
Unfortunately, since neither $g$ nor $\Lambda$ is known, this equation has no
predictive power.  It is comforting, however, that a reasonable choice of the
parameters gives a reasonable result.  For example, for $g = 1$ and 
$\Lambda = 1$ GeV, Eq. (\ref{spin}) yields a spin content of 0.57, which is
within ${\cal O}(\epsilon^2)$ of the experimental value \cite{spincontent} of
$0.27 \pm 0.11$ .  
In this case the expansion parameter $\epsilon$ is rather large, $\epsilon 
\approx 0.75$, so the effects of chiral loops are estimated to be very
important.

\section*{Acknowledgements}

It is a pleasure to acknowledge many helpful discussions with M. Suzuki and S.
Johnson.  I also thank M. Luty and J. March-Russell for patiently explaining
their work to me.  This work was supported in part by the Director, 
Office of Energy
Research, Office of High Energy and Nuclear Physics, Division of High Energy
Physics of the U.S. Department of Energy under Contract DE-AC03-76SF00098 and in
part by the Department of Education Graduate Assistance in Areas of National
Need Program.

\pagebreak

\section*{Figure Captions}

\vskip 1cm

\verb| |

{\bf Fig. 1} The one loop vertex renormalization (a) and wavefunction
renormalization (b).  The ``x'' represents the axial current operator.
\vskip 1cm

{\bf Fig. 2} Two loop vertex renormalization diagrams.
\vskip 1cm

{\bf Fig. 3} Some of the Feynman rules for baryon-meson interactions.
\vskip 1cm

{\bf Fig. 4} Two loop wavefunction renormalization diagrams.
\vskip 1cm

{\bf Fig. 5} The renormalization of the baryon axial current, viewed at the
quark level.  Diagram (a) = diagram (b), and (c) = (d).  Only (e) ends up
contributing to the overall renormalization.
\vskip 1cm

{\bf Fig. 6} Additional one and two loop diagrams that contribute to the
renormalization of the axial current.


\begin{thebibliography}{99}

\bibitem{old} J. Bijnens, H. Sonoda, and M. Wise, Nucl. Phys. B 261 (1985)
185. \\
J. Gasser, M.E. Sainio, and A. Svarc, Nucl. Phys. B 307 (1988)
779. \\
A. Krause, Helv. Phys. Acta 63 (1990) 1. \\
S. Weinberg, Phys. Lett. B 251 (1990) 288.

\bibitem{jm1} E. Jenkins and A. Manohar, Phys. Lett. B 255 (1991) 558.

\bibitem{jm2} E. Jenkins and A. Manohar, Phys. Lett. B 259 (1991) 353.

\bibitem{jenkins} E. Jenkins, Nucl. Phys. B 368 (1992) 190; \\
E. Jenkins and A. Manohar, Effective Field Theories of the Standard Model, ed.
U.-G. Meissner (World Scientific, Singapore, 1992).

\bibitem{bigm} H. Georgi, Phys. Lett. B 240 (1990) 447.

\bibitem{witten} E. Witten, Nucl. Phys. B 160 (1979) 57.

\bibitem{gervais} J. Gervais and B. Sakita, Phys. Rev. Lett.  52 (1984) 87; \\
Phys. Rev. D 30 (1985) 1795.

\bibitem{sd1} R. Dashen and A. Manohar, Phys. Lett. B 315 (1993) 425, 438.

\bibitem{sd2} E. Jenkins, Phys. Lett. B 315 (1993) 431, 441, 447.

\bibitem{djm1} R. Dashen, E. Jenkins, and A. Manohar, Phys. Rev. D 49 (1993) 
4713.

\bibitem{harvard} C. Carone, H. Georgi, and S. Osofsky, Phys. Lett. B 322 
(1994) 227.

\bibitem{lbl1} M. Luty and J. March-Russell, Nucl. Phys. B 426 (1994) 71.

\bibitem{lbl2} M. Luty, Phys. Rev. D 51 (1995) 2322

\bibitem{lbl3} M. Luty, J. March-Russell, and M. White, 
Phys. Rev. D 51 (1995) 2332.

\bibitem{djm2} R. Dashen, E. Jenkins, and A. Manohar, Phys. Rev. D 51 (1995)
3697

\bibitem{georgimanohar} A. Manohar and H. Georgi, Nucl. Phys. B 234 (1984) 189.

\bibitem{bluegeorgi} H. Georgi, Weak Interactions and Modern Particle Theory
(Addison-Wesley, Redwood City, 1984).

\bibitem{spincontent} J. Ellis and M. Karliner, 
Phys. Lett. B 313 (1993) 131; \\
Phys. Lett. B 341 (1995) 341

\end{thebibliography}
\end{document}